# Current-driven skyrmion motion along disordered magnetic tracks

Victor Raposo,[1] Ricardo Francisco Luis Martinez,[1] and Eduardo Martinez[1, a)]

[1]*Departamento de Fisica Aplicada. University of Salamanca, Salamanca, E-37008, Spain*

The motion of skyrmions along ferromagnetic strips driven by current pulses is theoretically analyzed by means of micromagnetic simulations. Analytical expressions describing the skyrmion dynamics during and after the current pulse are obtained from an extended rigid skyrmion model, and its predictions are compared with full micromagnetic simulations for perfect samples with a remarkable agreement. The dynamics along realistic samples with random disorder is also studied by both models. Our analysis describes the relevant ingredients behind the current-driven skyrmion dynamics, and it is expected to be useful to understand recent and future experimental.

**I. INTRODUCTION**

Multilayers consisting in a thin ferromagnetic (FM) strip sandwiched between a heavy metal (HM) and an oxide, or between two different heavy metals, are promising systems to develop novel magnetic memory devices such as the race-track[1]. These multilayers exhibit high perpendicular magnetocristalline anisotropy (PMA) and the lack of structural symmetry generates an interfacial Dzyaloshinskii-Moriya interaction (iDMI)[2,3] which promotes chiral magnetization patterns such as chiral domain wall (DWs)[4-8] or skyrmions (Sks)[9-11]. These textures can be driven by application of electrical current as due to the spin Hall effect (SHE)[5-8,11].

Nowadays, Sks are the focus of active research because they offer great potential as information carriers in robust, high-density, and energy-efficient spintronic devices. Bloch-like skyrmions have been observed in some ferromagnets without inversion symmetry[12,13,14] under the presence of an out of plane field ($H_z$). Their magnetization is antiparallel to the $H_z$ at their center and parallel to $H_z$ at the periphery, and it points along the azimuthal direction ($\vec{u}_\phi$) in the transition region[15,16]. The dynamics of Bloch-like Sk under direct currents driven by the adiabatic and non-adiabatic spin transfer torques has been studied theoretically by several works[15,16,17,18]. On the other hand, recent experiments[10] have shown that Néel-like Sks can be stabilized at room temperature and zero magnetic field in ultrathin FM/HM films deposited by sputtering, which makes their use appealing for the study of skyrmion structure and dynamics. Differently from Bloch-like Sks [12-18], the magnetization of these chiral Néel Sks points radially ($\vec{u}_\rho$) in the transition region between the inner and the outer parts [10,18] and its size can be tuned by the nature and the thickness of the materials that comprises the multilayers. Moreover, the current-driven dynamics of Néel-like is driven due to the SHE by short current pulses at speeds exceeding 100 m/s[11]. These observations promise an industrial integration but several challenges must be addressed before these objects can be integrated into spintronic devices. For instance, due to the sputtering, the pinning that each Sk experiences is random and local. Consequently, adjacent Sks can move with different velocities[19], and they can even collapse when one Sk is highly pinned at a strong local defect[11]. Therefore, a study of the motion of Néel Sk by current pulses under realistic conditions is timely and demanding.

Here, we theoretically study the motion of a single Néel Sk under current pulses along a perfect FM by means of micromagnetic simulations ($\mu M$). The current-driven Skyrmion dynamics (CDSD) is described analytically in the framework of the Thiele model[20], which considers the Sk as a rigid object (Rigid Skyrmion Model, RSM). Realistic FM tracks, where the material parameters depict a random dispersion in the form of grains, are also evaluated. The RSM is extended to take into account the pinning generated by the random grains, and a good quantitative agreement with the $\mu M$ results is demonstrated. This study provides a simple framework to describe recent and further experiments, and it will be useful to control and develop skyrmion-based devices.

## II. SKYRMION EQUILIBRIUM STATE AND MODELS.

A FM strip with a width of $w = 128\ nm$ and thickness of $t_{FM} = 1\ nm$ is studied. Typical parameters for a HM/FM/Oxide multilayer with strong iDMI are considered[5,8,10,11]: saturation magnetization $M_s = 10^6\ A/m$, exchange constant $A = 20\ pJ/m$, uniaxial magnetocristalline anisotropy $K_u = 0.8 \times 10^6\ J/m^3$ and iDMI parameter $D = 1.8\ mJ/m^2$. Fig. 1(a) shows the equilibrium state of a Néel-like Sk, which was obtained by minimizing the total energy of the system using Mumax.[21] A 2D grid with $2\ nm$ side cells was adopted. To extract the skyrmion size, the magnetization profile is described by a 360° DW profile[22,23]: $\vec{m}(\rho) = \sin\theta(\rho)\vec{u}_\rho + \cos\theta(\rho)\vec{u}_z$, with $\theta(\rho) = \theta_{DW}(\rho - R_{Sk}) + \theta_{DW}(\rho + R_{Sk})$, where $\theta_{DW}(\rho) = 2\operatorname{atan}[exp(\rho/\Delta)]$; $R_{Sk}$ is the skyrmion radius, $\Delta$ the DW width, $\rho$ is the radial coordinate from the Sk center, and $\vec{u}_\rho$ the radial unit vector in cylindrical coordinates. The micromagnetically computed components $m_x(x)$ and $m_z(x)$ of the Néel skyrmion along the $x$-axis ($\vec{u}_\rho = \vec{u}_x$, with $\rho = x$) are shown in Fig. 1(b) by open symbols. Solid lines depict the fit to the 360° DW profile. From the fit, the skyrmion radius and the DW width parameters are extracted: $R_{Sk} = 14\ nm$ and $\Delta = 16.4$ nm.

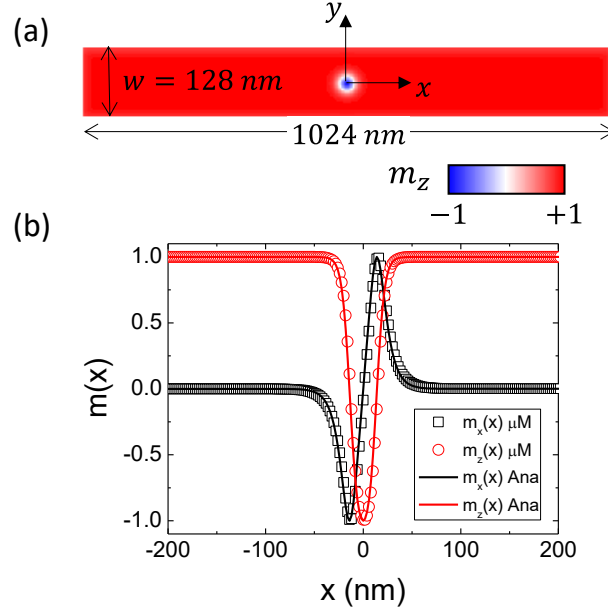

FIG. 1. (a) Micromagnetically computed skyrmion at rest. (b) Magnetization profile along the $x$-axis. Dots are $\mu M$ results and lines are fits to the 360° DW profile.

Within the micromagnetic model ($\mu M$), the magnetization dynamics is governed by the Gilbert equation augmented by the Slonczewskii-like spin-orbit torque (SL-SOT)[5,8,19,24]

$$\frac{d\vec{m}}{dt} = -\gamma_0 \vec{m} \times \vec{H}_{eff} + \alpha \vec{m} \times \frac{d\vec{m}}{dt} - \gamma_0 H_{SH}^0 \ \vec{m} \times (\vec{m} \times \vec{\sigma}), \qquad (1)$$

where $\gamma_0$ denotes the gyromagnetic ratio and $\alpha$ the Gilbert damping constant ($\alpha = 0.3$) respectively. $\vec{H}_{eff}$ is the deterministic effective field which includes the exchange, the magnetostatic, the uniaxial anisotropy and the DMI. The last term in eq. (1) is the SL-SOT due to the SHE,[5] where $H_{SH}^0 = \frac{\hbar \theta_{SH} J}{2\mu_0 |e| M_s t_{FM}}$, $\hbar$ is the Planck constant, $|e|$ is the electric charge, $\mu_0$ is the permeability of the free space, and $\theta_{SH}$ is the spin Hall angle ($\theta_{SH} = -0.33$). $J$ is the amplitude of the density current ($\vec{J} = J(t)\vec{u}_x$) and $\vec{\sigma} = \vec{u}_z \times \vec{u}_x = \vec{u}_y$ is the unit vector of the spin current generated by the SHE in the HM.

The CDSD can be also described with the formalism introduced by Thiele[18,20,24]:

$$\vec{G} \times \vec{v} + \alpha \mathcal{D} \vec{v} = \vec{F}_{SHE} + \vec{F}_C, \qquad (2)$$

where $\vec{v} = \frac{d\vec{r}}{dt} = \frac{dX}{dt}\vec{u}_x + \frac{dY}{dt}\vec{u}_y$ is the skyrmion velocity. $\vec{G} = -4\pi(pq)\frac{\mu_0 M_s}{\gamma_0} t_{FM}\vec{u}_z$ is the gyrovector, where $p$ and $q$ represent the skyrmion polarity and topological charge respectively. Here we discuss the case of a left-handed Néel skyrmion, as represented in Fig. 1(a)-(b), where $p = -1$ and $q = +1$. $\mathcal{D}$ is the 2 × 2 diagonal dissipation matrix, with elements $D_{ij} = D\delta_{ij} \approx -\frac{\mu_0 M_s}{\gamma_0} t_{FM}\left(\frac{\pi^3 R_{Sk}}{\Delta}\right)$. $\vec{F}_C$ is the restoring force due to the confinement (*i.e.* by the strip edges), which in a first approximation can be described as $\vec{F}_C = -kY\vec{u}_y$ with $k$ being an elastic constant, which is estimated from a single $\mu M$ simulation[25]. $\vec{F}_{SHE}$ is the driving force due to the SHE, which is given by $\vec{F}_{SHE} \approx -\frac{\hbar\theta_{SH}J}{2e}(pq)\pi^2\eta R_{Sk}\vec{u}_x$. Eq. (2) describes the motion of a rigid skyrmion with a characteristic size of given by $\eta R_{Sk}$[25].

## IV. SKYRMION DYNAMICS UNDER CURRENT PULSES ALONG IDEAL STRIPS

The CDSD under current pulses with zero rise and falling time is evaluated both by $\mu M$ and RSM models. The temporal evolution of the longitudinal $X(t)$ ($x$-direction) and the transverse $Y(t)$ ($y$-direction) displacements are shown in Fig. 2(a) and (b) for pulses with fixed length of $t_p = 5\ ns$ and different amplitudes $J$. Both $X(t)$ and $Y(t)$ increase during the pulse ($t \leq t_p$). Once the pulse is switched off ($t > t_p$), $Y(t > t_p)$ starts to decrease returning to zero for sufficient long times. On the contrary, $X(t > t_p)$ continues forward when the pulse is turned off, and it tends asymptotically to a terminal value $X_F$.

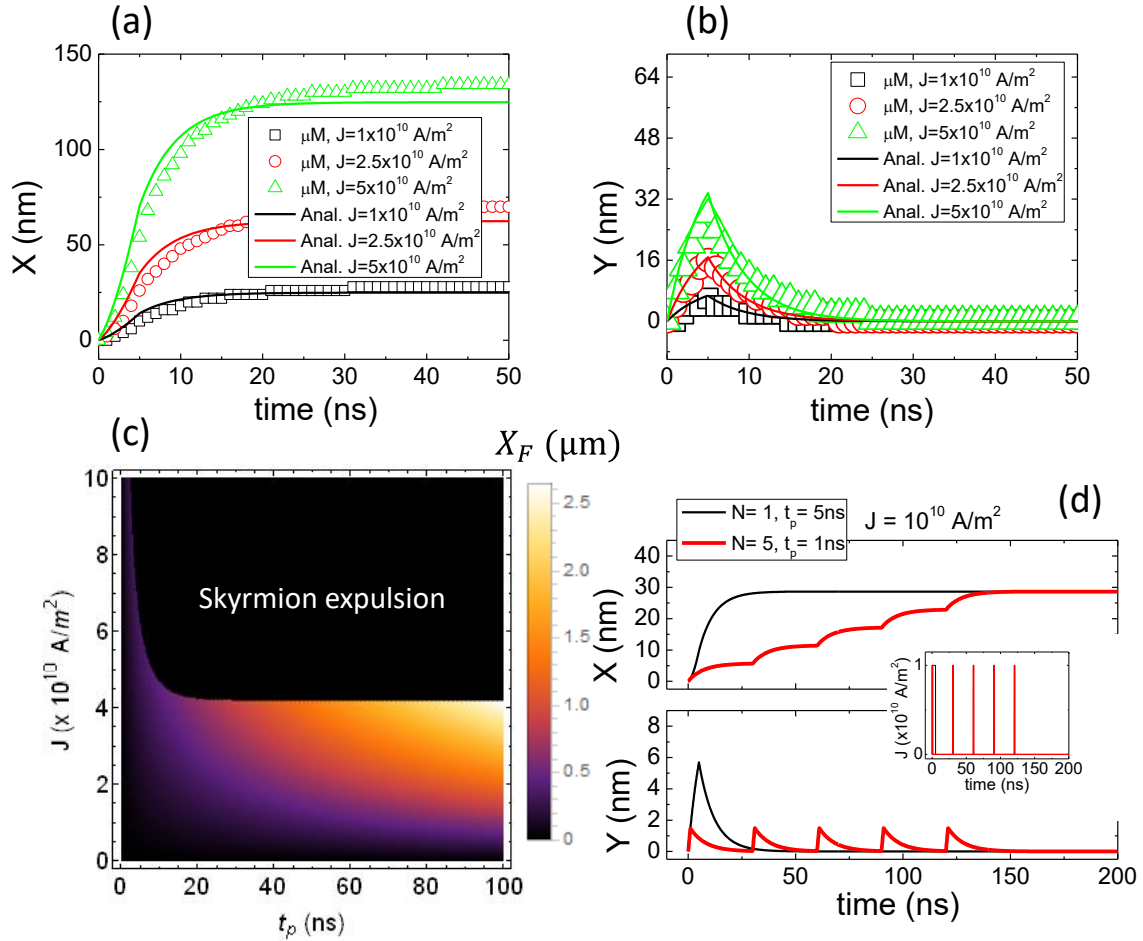

FIG. 2. CDSD under current pulses with different amplitudes ($J$) and fixed length ($t_p = 5\ ns$) along a perfect strip: (a) and (b) show $X = X(t)$ and $Y = Y(t)$ as a function of time respectively. Dots are $\mu M$ results and lines are analytical RSM predictions. (c) $X_F$ as a function of the amplitude $J$ and the length $t_p$ of the pulse. The skyrmion is assumed to be expelled from the strip if $Y(t) > Y_{th} = 50\ nm$ (black top area). (d) $X = X(t)$ and $Y = Y(t)$ for a single pulse ($N = 1$) with $J = 10^{10} A/m^2$ and $t_p = 5\ ns$ (black line) and for five pulses ($N = 5$) with $t_p = 1\ ns$ and the same amplitude (red line).

The CDSD under a current pulse can be also analytically described by the RSM, eq. (2). Its analytical solution during the current pulse ($t \leq t_p$) is:

$$X(t \leq t_p) = \frac{F_{SHE}}{\alpha D}\left[t + \frac{G^2}{\alpha D\, k}\left(-1 + \exp\left(-\frac{t}{\tau}\right)\right)\right], \quad (3)$$

$$Y(t \leq t_p) = \frac{G\, F_{SHE}}{\alpha D\, k}\left[\left(1 - \exp\left(-\frac{t}{\tau}\right)\right)\right], \quad (4)$$

where $\tau$ is the characteristic relaxation time given by $\tau = \left|\frac{G^2 + (\alpha D)^2}{\alpha D\, k}\right|$. The post-pulse ($t > t_p$) dynamics is determined by:

$$X(t > t_p) = X(t_p) + \frac{G}{\alpha D} Y(t_p)\left(1 - \exp\left(-\frac{t - t_p}{\tau}\right)\right), \quad (5)$$

$$Y(t > t_p) = Y(t_p) \exp\left(-\frac{t - t_p}{\tau}\right), \quad (6)$$

where $X(t_p)$ and $Y(t_p)$ represent the skyrmion position at the end of the pulse ($t = t_p$) as obtained from eqs. (3) and (4). These analytical predictions (solid lines) are compared to the $\mu M$ results (open symbols) in Fig. 2(a) and (b). A remarkable agreement is observed between both models, which allows us to predict some interesting points. In particular, we have studied how the terminal longitudinal displacement ($X_F \equiv X(t \gg t_p)$) depends on both the amplitude ($J$) and the duration ($t_p$) of a single current pulse. It has to be taken into account that $X_F$ scales with the transverse displacement at the end of the pulse $Y(t_p)$. If during the pulse the skyrmion approaches to the edge ($Y(t) \to \frac{w}{2}$), the repulsion from it cannot balance the transverse pushing force due to the current, and consequently, the skyrmion is expelled from the strip[18,19,24]. This imposes a limit in the maximum transverse displacement ($Y_{th}$) which can be reached without skyrmion annihilation. $X_F$ as function of $J$ and $t_p$ can be seen in Fig. 2(c), where it was assumed that the skyrmion annihilates if $Y_{th} \geq 50\ nm$ (black area on top in Fig. 2(c)). See Ref. 26 for an analytical estimation of the product $(J\, t_p)$ above which the Sk reaches the threshold transverse value $Y_{th}$. These results can be inferred from eqs. (3)-(6): the terminal longitudinal distance $X_F$ without expulsion is $X_F = \frac{F_{SHE} t_p}{\alpha D}$, which is proportional to the product $(J\, t_p)$. Therefore, the same $X_F$ can be achieved under different combinations of $J$ and $t_p$, as it clearly seen in Fig. 2(c).

Other interesting observation is that the same $X_F$ can be achieved with a single pulse of length $t_p = T$ or with $N$ pulses each one with $t_p = \frac{T}{N}$. This is shown in Fig. 2(d), where the displacement for a single pulse with $(J, t_p = T) = (10^{10}\ A/m^2, 5\ ns)$ is compared to the one achieved by a train of $N = 5$ pulses of the same $J$ but length $t_p = \frac{T}{N} = 1\ ns$, when they are applied every $30\ ns$. This is interesting for applications, because short pulses are required to minimize unwanted Joule heating effects due to the current injection[27,28].

## V. SKYRMION DYNAMICS UNDER CURRENT PULSES ALONG REALISTIC STRIPS

Former analysis was performed under ideal conditions assuming a perfect strip. However, real samples present unavoidably imperfections. In order to evaluate realistic conditions, the disorder is taken into account in the $\mu M$ by considering that the FM strip consists on grains[29]. We assume the easy axis anisotropy direction ($\vec{u}_K$) is distributed among a length scale defined by a characteristic grain size of $50\ nm$. $\vec{u}_K$ of each grain is mainly directed along the perpendicular direction ($z$-axis) but with a small in-plane component, which is randomly generated over the grains. The maximum percentage of the in-plane component of the uniaxial anisotropy unit vector is 5%. A typical grain pattern (GP) is shown in Fig. 3(c). Four different grain patterns have been evaluated to obtain statistic results. $\mu M$ results of the temporal evolution of $X(t)$ under a train of $N = 5$ pulses and $t_p = 6\ ns$ are shown in Fig. 3(a)-(c) for different $J$. Here, the time between consecutive pulses is equal to $t_p$. Each thin black line corresponds to one of the four GPs evaluated. The presence of disorder imposes a threshold pinning current density $J_p \sim 1.5 \times 10^{10}\ A/m^2$ below which the skyrmion is hardly displaced from its initial location (Fig. 3(a)). Note that the Sk stops at $\sim 20\ ns$, well before the end of the train of pulses $t = 2Nt_p = 60\ ns$. On the contrary, the Sk is significantly driven by the current pulses for $J > J_p$ (Fig. 3(b)-(c)).

These $\mu M$ results can be also described by the RSM, where the pinning can be accounted by adding a pinning force $\vec{F}_p(X,Y) = -\nabla V_p$ to Eq. (2). $V_p(X,Y)$ is a 2D pinning potential given by $V_p(X,Y) = V_0 \cos\left(\frac{2\pi}{\lambda}X\right)\cos\left(\frac{2\pi}{\lambda}Y\right)$, with $V_0$ being the energy barrier and $\lambda$ is parameter related to the spatial periodicity. $V_0$ is related to the critical depinning current $J_p$ as $V_0 = \frac{\lambda}{2\pi}F_p$, with $F_p = \frac{\hbar\theta_{SH}J_p}{2e}(pq)\pi^2\eta R_{Sk}\vec{u}_x$. A snapshot of $V_p/V_0(X,Y)$ is shown in Fig. 3(d) for $\lambda = 15\ nm$. Solid red lines in Fig. 3(a)-(c) depict $X(t)$ vs $t$ as computed by the RSM. The average velocity, computed as the terminal displacement after $100\ ns$ ($X(100\ ns)$) divided by the total integrated time ($5t_p = 30\ ns$), is compared to $\mu M$ results in Fig. 3(f). A good quantitative agreement is observed between both models, indicating that the extended RSM constitutes a simple framework to describe $\mu M$ results and also experimental observations.

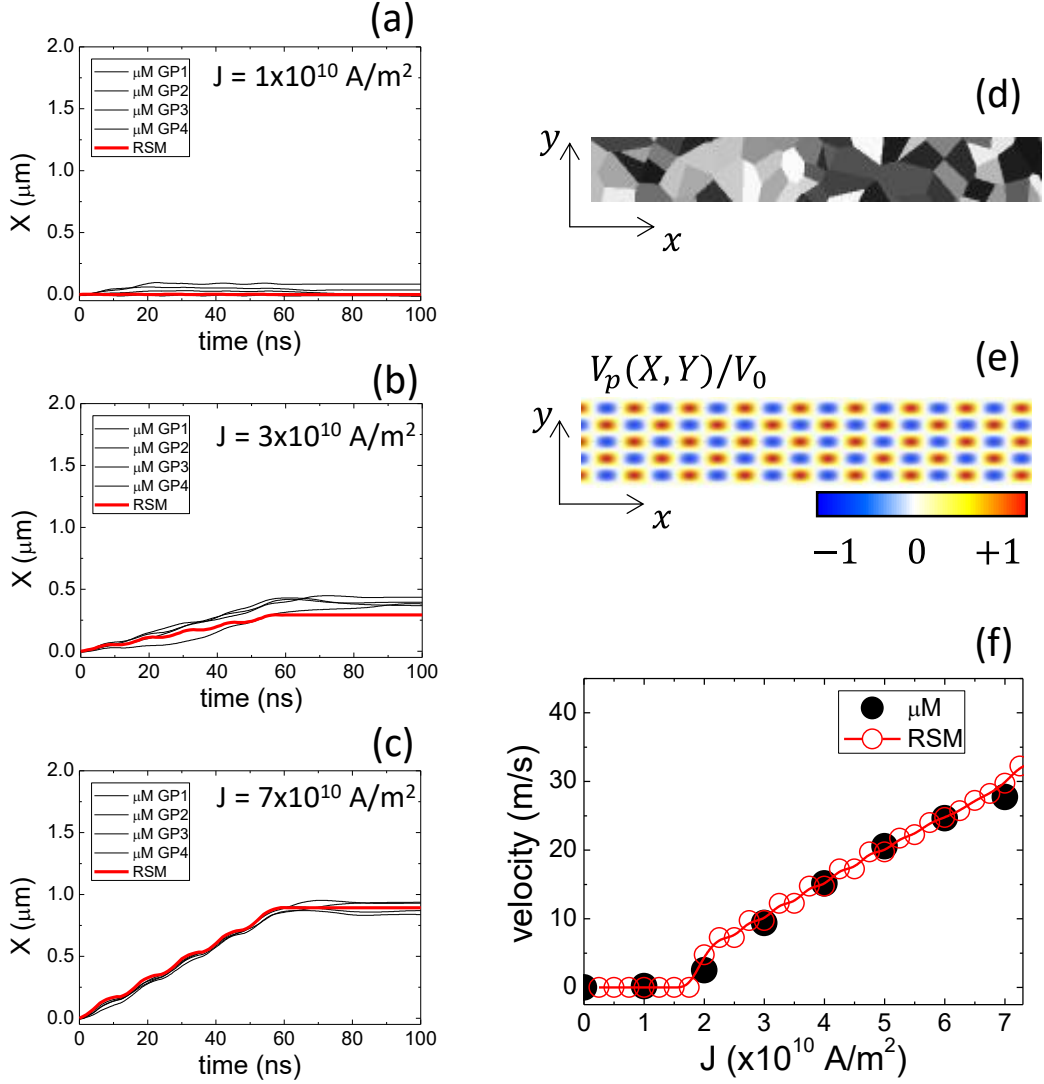

FIG. 3. CDSD under current pulses along a realistic strip. (a)-(c) show $X$ vs $t$ for three different values of $J$. Five consecutive pulses with $t_p = 5\ ns$ are applied. Time between pulses is equal to $t_p$. Thin black lines are $\mu M$ results and the thick red lines were obtained from the RSM using $J_p = 1.0 \times 10^{10}\ A/m^2$ and $\lambda = 15\ nm$. (d) A typical grain pattern considered in the micromagnetic simulations. (e) Normalized pinning potential $V_p/V_0$ adopted in the RSM as a function of position. (f) Average skyrmion velocity ($X(100\ ns)/(5t_p)$) as a function of $J$.

## VI. CONCLUSIONS

The CDSD has been studied by $\mu M$. The results can be analytically described by a RSM, which allows us to obtain the relevant parameters governing the dynamics. In particular, the acceleration and deceleration under current pulses is determined by the same characteristic time $\tau$, which scales with the inverse of the damping of the system ($\tau \sim 1/\alpha$). Our analytical description indicates that the terminal longitudinal displacement $X_F$ under current pulses scales with the pulse length $t_p$, and the same $X_F$ can be a achieved with a single pulse of a given duration $T = t_p$ or with $N$ consecutive pulses of the same amplitude $J$ of with length $t_p = T/N$. These observations are interesting for applications to minimize the unwanted Joule heating effect, which scales with $J^2$ and $t_p$. $\mu M$ results under realistic conditions, including disorder, can be also described by the RSM. This study will be useful to understand experimental measurements and consequently, to control and develop skyrmion-based devices.

**ACKNOWLEDGMENTS**

This work was supported by project WALL, FP7-PEOPLE-2013-ITN 608031 from the European Commission, project MAT2014-52477-C5-4-P from the Spanish government and project SA282U14 from the Junta de Castilla y Leon.

**25** From eq. (3) and (4), the steady longitudinal velocity $V_{x,st}$ and the terminal transverse displacement $Y_{st}$ can be obtained in the limit of $t \gg \tau$: $V_{x,st} = \frac{F_{SHE}}{\alpha D}$ and $Y_{st} = \frac{G\,F_{SHE}}{\alpha D\,k}$. From these expressions, the inputs $(b, k)$ for the Rigid Skyrmion Model (RSM) can be obtained from a single $\mu M$ simulation. For example, the $\mu M$ results for the terminal longitudinal velocity and transverse displacement under a current of $J = 0.5 \times 10^{10}\ A/m^2$ are $V_{x,st} \approx 3\ m/s$ and $Y_{st} \approx 6\ nm$ respectively. From these $\mu M$ data, we obtain $k \approx 3.6 \times 10^{-5} N/m$ and $\eta \approx 1.8$ using $R_{Sk} = 14$ nm and $\Delta = 16.4$ nm. These inputs were adopted to obtain the RSM results in the text.

**26** The skyrmion is expelled from the strip if $Y(t) \gtrsim Y_{th} \equiv \frac{w}{2}$. The combination of pulse amplitudes $J$ and pulse lengths $t_p$ satisfying this expulsion condition is obtained from eq. (4), and represented in Fig. 2(c). Note that the skyrmion can be expulsed before the end of the pulse at $t < t_p$. In order to provide an analytical expression for this condition, the exponential within the brackets in eq. (4) is expanded up to first order, and the following condition is deduced: $J\,t_p \geq (w|e|\tau \alpha k)/(4\Delta \eta \hbar \theta_{SH})$ where $k$ the elastic constant respectively, $w$ is the width of the strip and $\tau$ is the characteristic time as given in the text. Note however, that this approximated expression underestimates the threshold skyrmion expulsion for pulses with $t_p \gtrsim 10\ ns$, and therefore, the combination of $J$ and $t_p$ resulting in expulsion is more accurate as directly obtained from eq. (4).

**27** H. Fangor, D. S. Chernyshenko, M. Franchin, T. Fischbaher and G. Meier. "Joule heating in nanowires" Phys. Rev. B. 84, 054437 (2011).

**28** S. Moretti, V. Raposo and E. Martinez. "Influence of Joule heating on current-induced domain wall depinning" J. Appl. Phys. 119, 213902 (2016).

**29** E. Martinez, O. Alejos and M. A. Hernandez, V. Raposo and L. Sanchez-Tejerina and S. Moretti. "Angular dependence of current-driven chiral walls" Appl. Phys. Express, 9, 063008 (2016).